\documentclass{article}
\usepackage{graphicx}
\graphicspath{%
    {converted_graphics/}
    {C:/Dropbox/1-kgl-top/Papers/2014/DualChamberConcept/}
}
\begin{document}

\begin{center}
{\LARGE A Dual Diffusion Chamber for}\vskip8pt

{\LARGE Observing Ice Crystal Growth on c-Axis Ice Needles}\vskip10pt

{\Large Kenneth G. Libbrecht}\vskip6pt

{\large Department of Physics, California Institute of Technology}\vskip1pt

{\large Pasadena, California 91125}\vskip-1pt

\vskip18pt

\hrule\vskip1pt \hrule\vskip14pt
\end{center}

\textbf{Abstract.} We describe a dual diffusion chamber for observing ice
crystal growth from water vapor in air as a function of temperature and
supersaturation. In the first diffusion chamber, thin c-axis ice needles with tip radii $\sim 100$ nm are grown to lengths of $\sim 2$ mm.
The needle crystals are then transported to a second diffusion chamber where
the temperature and supersaturation can be independently controlled. By
creating a linear temperature gradient in the second chamber, convection
currents are suppressed and the supersaturation can be modeled with high
accuracy. The c-axis needle crystals provide a unique starting geometry
compared with other experiments, and the dual diffusion chamber allows rapid
quantitative observations of ice growth behavior over a wide range of
environmental conditions.

\section{Introduction}

The basic physics underlying crystal growth is remarkably rich, as processes
over many length scales -- from short-range molecular dynamics to long-range
diffusion effects -- work in concert to determine surface growth rates and
crystal growth morphologies. The formation of faceted dendritic structures
has been particularly challenging to understand in detail, requiring
simultaneous descriptions of both the surface attachment kinetics
(responsible for faceting) as well as particle and/or heat transport via
diffusion surrounding the crystal (driving dendritic growth). Even realizing
stable computer simulations that qualitatively reproduce observed structures
has been achieved only recently \cite{griffeathgravner, barrettgarcke}.

The formation of snow crystals -- ice crystals grown from water vapor in air
-- is the most-studied example of faceted dendritic crystal growth to date,
yet our understanding of these common structures is surprisingly poor. The
dramatic variation of snow crystal morphology with growth temperature has
been known for over 75 years, for example, yet this seemingly simple
behavior is still without a satisfactory explanation \cite{libbrechtreview}.
One reason for this state of affairs is that the surface attachment kinetics
depend in detail on the molecular structure of the ice surface, including
surface melting, which is itself not well understood. In addition, we have
recently found that the attachment kinetics depend on the mesoscale surface
structure, further complicating our picture of ice crystal growth \cite%
{comprehensive}.

Making significant progress in crystal growth physics often requires
quantitative comparisons between experimental observations and comprehensive
computer modeling that incorporates all relevant physical processes. For the
case of ice growth from water vapor, this begins with precise measurements
of crystal growth rates and morphologies under well-defined conditions. Many
of the ice-growth studies reported in the literature are not well-suited for
quantitative analysis \cite{critical}, and there remains much room for
improvement on the experimental front.

Focusing on ice growth from water vapor, a variety of experimental methods
have been used for measuring growth rates and morphologies under different
conditions. We begin with a brief summary of these methods.

\textbf{Growth on a flat substrate.} In this method one begins with a small
seed crystal resting on a flat substrate, and then observes its subsequent
growth as a function of temperature, supersaturation, background gas
pressure, and other conditions. This method affords many advantages,
including: 1) small crystals can be observed, which is important for
minimizing diffusion effects and measuring surface attachment kinetics; 2)
interferometric techniques can be used for making precise measurements of
growth rates; 3) aligning a crystal facet to the substrate gives a
well-defined growth morphology relative to the substrate, and 4) the
supersaturation is set by the temperature difference between the growing
crystal and a nearby water vapor reservoir, which can yield exceptionally
good determinations of this important quantity. The substrate method was
used to make the most accurate measurements to date of the attachment
coefficient for ice growth over a range of conditions \cite{growthmeas}, as
well as exploring aspects of structure-dependent attachment kinetics \cite%
{comprehensive}. Overall this method is perhaps the most useful for making
precise measurements of ice growth rates in well-defined conditions.

Disadvantages of this method include:\ 1) ice/substrate interactions can
substantially alter the ice growth rates \cite{systematicerrors}; 2) the
supersaturation must remain below $\sigma _{water}$, limited by the
formation of droplets on the substrate; and 3) neighboring crystals (near
the test crystal but perhaps unseen) can strongly affect the
supersaturation, so care must be taken to isolate a single test crystal.

\textbf{Growth on a filament substrate.} This method dates back to Nakaya's
measurements in the 1930s using rabbit hair \cite{nakaya}, and more recent
examples include the use of thin glass capillaries \cite{knight5,
knightnelson}. Advantages of this method include:\ 1) the thin filament
reduces substrate interactions somewhat and allows observations of full
three-dimensional growth morphologies; 2) supersaturations above $\sigma
_{water}$ can be obtained if the growing crystal shields the filament to
prevent droplet formation, and 3) single crystals can be observed, allowing
good determinations of the surrounding supersaturation.

Disadvantages of this method include: 1) substrate interactions are
significant and difficult to control; 2) the crystal morphology is typically
not aligned with respect to the substrate, complicating the analysis of
substrate interactions; 3) observing very small crystals is usually
impractical, making it difficult to reduce diffusion effects. While the
filament technique has been particularly useful in qualitative studies of
ice growth, its disadvantages are nontrivial, especially substrate
interactions, making it difficult to realize precise measurements. 

\textbf{Growth in free fall.} In this method one typically nucleates
crystals in air, allows them to grow in free-fall, and then observes them as
they fall onto a substrate at the bottom of the growth chamber \cite{gonda,
gonda2}. Advantages of this method include: 1) there are no substrate
interactions to contend with; 2) large numbers of crystals can be observed,
allowing the study of rare morphologies (e.g. triangular plate crystals \cite%
{triangular}); 3) very small crystals can be studied; and 4) free-fall
growth chambers typically have larger volume/surface ratios than other
growth chambers, allowing relatively cleaner environments (with fewer
contaminants in the air) to be used.

Disadvantages include: 1) the presence of many growing crystals makes it
difficult to determine the supersaturation accurately; 2) growth in
near-vacuum conditions is typically not practical; 3) the growth cannot be
observed continuously, and 4) the conditions are typically not uniform
inside the growth chamber.

The free-fall method is especially useful for performing morphological
studies with large numbers of crystals, but again this method is not well
suited for making precise growth measurements, mainly because the
supersaturation is difficult to determine.

\textbf{Levitated crystals.} Observing individual levitated crystals is a
particularly appealing experimental method, as in principle it has a great
many advantages, including: 1) the complete absence of substrate
interactions; 2) very small crystals can be used; and 3) the crystal growth
can be continuously monitored. In many ways levitated crystals are ideally
suited for making precise growth measurements.

Disadvantages include:\ 1) the technical challenges are much greater than
with the other methods; 2) maintaining levitation as the crystal mass
increases can be problematic, and 3) controlling and determining the
supersaturation may be difficult if ice condensation on the levitation
hardware is close to the growing crystals.

The growth of ice crystals in ion traps has been reported \cite{levitated},
and ice crystals held in air flows have also been observed \cite{fukuta}.
Levitation of ice crystals by thermo- and photophoretic forces has been
demonstrated as well \cite{levitatetherm}, as has acoustic levitation of ice
particles \cite{acoustic}. With the exception of \cite{fukuta}, these
techniques have so far remained mainly in the demonstration phases, and have
not yet been used to produce comprehensive measurements of ice growth rates.

\textbf{Growth on ice needles.} In this paper we describe the development of
a novel method for observing the growth of ice crystals at the ends of
slender ice needles. The needles are grown near $-6$ C at high
supersaturation by applying 2 kilovolts DC to a wire substrate. Once formed,
the needles are transported to a second chamber for subsequent measurements.
Advantages of this technique include: 1) there are no substrate
interactions, even though the crystal is held in a stationary position; 2)
the growth can be continuously monitored; 3) high supersaturations can be
used to grow complex morphologies; and 4) a rapid turnaround makes it
possible to study many crystals quickly.

The method has a number of disadvantages also, including: 1) the hardware is
technically complex (although in practice quite reliable); 2) the ice needle
interferes with the crystal growth at the tip (although in a well-defined
way); 3) the supersaturation is somewhat influenced by the underlying wire
substrate and the presence of neighboring needles near the test crystal.

We believe that the dual-chamber apparatus described here is well-suited for
the study of fast-growing morphologies like thin plates, stellar crystals,
and hollow columns under controlled conditions in air. Our hope is that
quantitative measurements of these morphologies will provide new insights
into Structure-Dependent Attachment Kinetics (SDAK) \cite{comprehensive}. It
appears that the SDAK mechanism is an important player in the growth of snow
crystals, particularly the SDAK instability, and a better understanding of
this enigmatic phenomenon is needed. Our dual-chamber apparatus should allow
a quantitative exploration of the SDAK instability over a range of
conditions, thus leading to a greater understanding of the detailed physical
processes underlying ice crystal growth dynamics.

\begin{figure}[ht] 
  \centering
  \includegraphics[width=5.8in,keepaspectratio]{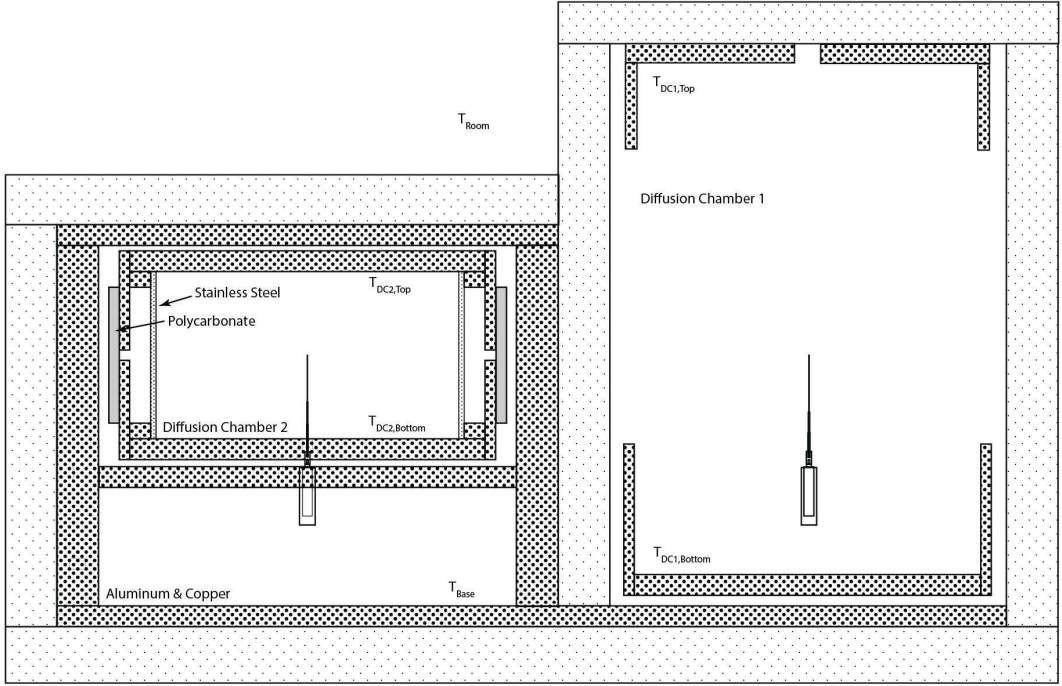}
  \caption{A schematic view of the
dual-chamber apparatus. Ice needle crystals are created in Diffusion Chamber
1 (DC1, on the right), on the tip of a wire near the center of the chamber.
The wire is mounted to a small gearhead motor (below the wire), which is in
turn mounted to a manipulator arm (not shown here). After the needles have
grown to lengths of $\sim 2$ mm, the manipulator arm translates the
motor, wire, and ice crystals horizontally to Diffusion Chamber 2 (DC2, on
the left). The growth conditions are better defined in the second chamber,
typically with a lower supersaturation than in DC1. A microscope objective
mounted near the back wall of DC2 (not shown here) allows the subsequent ice
growth to be observed.}
  \label{DCschematic1}
\end{figure}

\section{Dual Chamber Design}

Figure \ref{DCschematic1} shows the overall layout of our dual-chamber
apparatus. The use of two side-by-side diffusion chambers allows each to be
optimized separately for its intended function. Diffusion Chamber 1 (DC1)
was designed with an exceptionally high supersaturation, near 100\% near $-6$
C, as this is needed for reliably producing c-axis ice needles. Diffusion
Chamber 2 (DC2) was designed to produce a well-known temperature and
supersaturation in the vicinity of the growing crystals, where both
parameters can be varied over a wide range. The two diffusion chambers are
built upon a common base plate that is cooled to $-35$ C using a
recirculating chiller.

Also common to both chambers is the wire tip upon which ice crystals
initially grow. At the base of the wire assembly is a small gearhead motor
(Faulhaber 06/1 64:1 gearhead with 0615 C 003 S motor) that is used to
rotate the wire, thus positioning the ice needles for observation in DC2.
Typically the wire is rotated so the c-axis of a particular ice needle is
perpendicular to the optical axis of the microscope. In this way the needle
diameter as a function of time can be measured, along with the tip crystal
diameter and thickness. Other observation angles yield additional
information about the crystal morphology, etc.

The wire itself is a sharpened stainless steel acupuncture needle (SEIRIN J
Type, 120 $\mu $m in diameter, 40 mm long). The wire is supported by
telescoping stainless steel tubes that connects it to the rotation motor. A
manipulator arm mounted below the motor translates the motor/needle assembly
back and forth between the two diffusion chambers. A pair of sliding
shutters isolate the diffusion chambers from each other, minimizing air
currents between the two, and both shutters are opened only briefly when the
motor/needle assembly is being transported between the two chambers.

\subsection{Diffusion Chamber 1}

DC1 was designed to be an electric ice needle factory, capable of reliably
producing c-axis ice needles on demand, following the procedures described
in \cite{eneedles}. The inner dimensions of the diffusion chamber are 250 mm
square by 350 mm in height. The walls are made from a combination of acrylic
and polycarbonate sheets, surrounded by styrofoam insulation. The styrofoam
is isolated from the inside of the chamber by the plastic sheets to avoid
outgassing from the styrofoam. Outgassing from the plastic appeared to be
negligible after some preliminary baking of the system.

A copper water reservoir is placed near the top of DC1, and this metal
assembly is typically run at a temperature of 59.0$\pm 0.4$ C. Metal side
extensions reduce condensation on the walls near the top of the chamber
(near the water reservoir) and serve to increase the temperature gradient
(and thus the supersaturation) near the center of the chamber.

The aluminum assembly near the bottom of DC1 (see Figure \ref{DCschematic1})
is typically held at a temperature of $-35.0\pm 0.3$ C, and again the side
extensions increase the temperature gradient near the center of the chamber.
The finite thickness of the insulating walls around DC1 mean that the
temperature profile within the chamber depends slightly on the room
temperature surrounding the apparatus. This effect was usually negligible,
as the room temperature was also regulated by the building air-handling
system.

\begin{figure}[t] 
  \centering
  \includegraphics[width=4in,keepaspectratio]{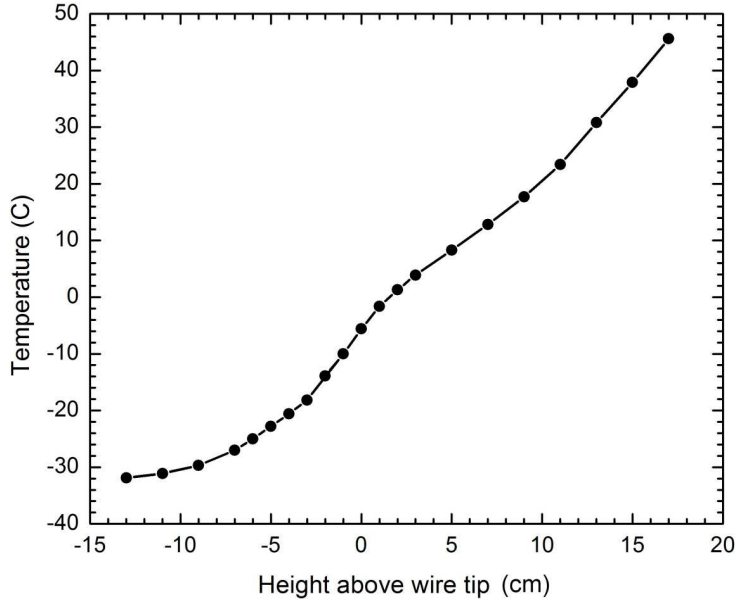}
  \caption{The measured DC1 temperature
profile during operation. The position measurement is relative to the wire
tip upon which the ice needles grow. The water reservoir at the top of DC1
is held at 59 C, while the base is held at $-35$ C.}
  \label{dc1temp}
\end{figure}

Figure \ref{dc1temp} shows the vertical temperature profile along the
central axis of DC1. This and other temperature measurements reveal that the
isothermal surfaces within the chamber are not horizontal, which in turn
means that there must be some convective air currents present. These slow
convective motions likely significantly affect the supersaturation within
the chamber, and modeling these effects would be challenging. However, the
sole purpose of DC1 is to produce a high supersaturation near $-6$ C at the
wire tip, as these are the conditions needed for producing c-axis ice
needles \cite{eneedles}. We use crystal growth rates as an indicator of the
supersaturation (see below), but its exact value is otherwise not determined.

\begin{figure}[t] 
  \centering
  \includegraphics[width=3.1in,keepaspectratio]{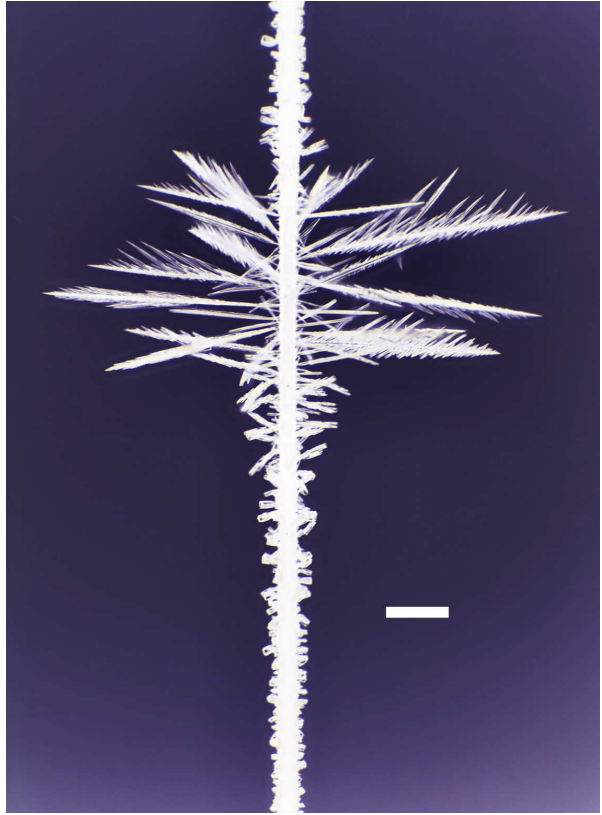}
  \caption{An image showing ice crystals
growing on a segment of 200-$\protect\mu $m-diameter nylon fishing line
hanging in the center of DC1. The scale bar is 1 mm long, and the crystal
growing time was 19 minutes. The crystal morphology is strongly temperature
dependent, with the long \textquotedblleft fishbone\textquotedblright\
crystals appearing near $-5$ C. The best wire tip location for producing
c-axis electric needles is just below the fishbone peak.}
  \label{needlecluster}
\end{figure}

Figure \ref{needlecluster} shows an example of ice crystals growing near a
temperature of $-6$ C within DC1, on a nylon filament placed along its
central axis. The crystal growth morphology is strongly temperature
dependent in this range, so observations like these provide an excellent
surrogate for \textit{in situ} temperature measurements as a function of
height. The filament is typically placed in DC1 at the beginning of each run
to verify that the system is performing satisfactorily, and it is otherwise
not present.

The fastest growing crystals seen in Figure \ref{needlecluster} are the
\textquotedblleft fishbone\textquotedblright\ crystals \cite{fishbones}
growing near $-5$ C. Through trial-and-error we determined that the best
location for creating c-axis electric ice needles is just below the fishbone
cluster. The temperature profile in DC1 was adjusted by changing the top and
bottom temperatures, along with the heights of both sets of metal side
extensions, so that the vertical location of the wire tip was near this
optimum temperature. Once the chamber reached its operating state, we found
that the vertical position of the needle cluster was stable to $\pm 0.25$
mm, which was the limit of our measurements.

The following procedure is followed to create a set of electric ice needles:
1) The wire is transferred to DC1, and ice crystals on the wire are knocked
off as much as possible using a cold metal cylinder. Occasionally, if the
wire is carrying a lot of ice, a warm metal cylinder is inserted through a
hole in the top of DC1 to melt and clean ice crystals from the wire.
Following this melt step, the cold cylinder is used to touch the wire and
freeze the remaining water film. 2) With the wire tip in place, one ml of
air containing a chemical additive is inserted into DC1 just below the wire.
Silicone caulk (GE Silicone II) vapor is the best additive we have found to
date, and this contains acetic acid vapor with an unknown mix of other
volatile compounds. 3) After inserting the chemical additive, 2 kV DC is
applied to the wire. Thin ice needles typically begin growing within a few
seconds after the voltage is applied, and their growth is monitored using a
long-distance microscope located outside DC1. 4) After the needles have
grown to lengths of roughly 2 mm, which takes 10-15 seconds, the high
voltage is turned off and a thin horizontal plate is immediately positioned
above the wire. This plate reduces the supersaturation so fishbone crystals
do not begin growing on the ice needles after the voltage is turned off. 5)
The wire assembly with ice needles is then transferred to DC2 for further
observations.

\begin{figure}[htb] 
  \centering
  \includegraphics[width=3.5in,keepaspectratio]{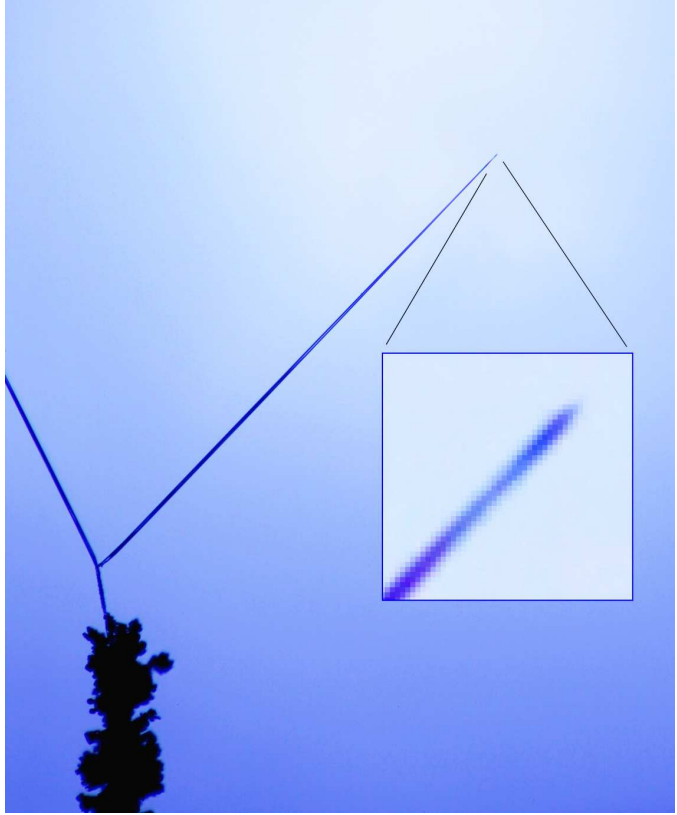}
  \caption{An example of a c-axis electric
ice needle. The total needle length is 2.3 mm, and the close-up inset box is
50 $\protect\mu $m on a side. Note that the substrate wire, seen in the
lower part of the image, is covered with frost crystals.}
  \label{eneedles}
\end{figure}

Figure \ref{eneedles} shows an example of an electric needle crystal
immediately after being transferred to DC2. The needles are quite robust and
usually survive the transfer well. Within roughly ten seconds one can
position the wire tip in DC2, rotate it to its desired position, focus on a
particular ice needle, and begin monitoring the subsequent growth of the
crystal.

Several conditions are required to successfully produce c-axis electric ice
needles \cite{eneedles}:\ 1) the temperature must be near $-6$ C; 2) the
supersaturation must be sufficiently high, above roughly 100 percent; 3) the
applied voltage must be roughly 1 kV or higher; and 4) some chemical
additive vapor must be present in the air. If all these conditions are met,
then typically several c-axis needles grow out from the needle tip in random
directions, and this happens with essentially a 100 percent yield. If any of
the conditions are not met, then either no electric needles appear, or more
commonly the needles will not grow along the c-axis.

When the nylon filament is placed in DC1, the fastest fishbone crystals grow
with tip velocities of 3.5-4.0 $\mu $m/sec. This observation serves as a
surrogate supersaturation measurement, since the tip growth velocity depends
approximately linearly on supersaturation in this range \cite{eneedles}. If
the wire is positioned in the chamber with no applied voltage, then the
fastest fishbone crystals grow with tip velocities of 7-8 $\mu $m/sec. The
factor of two increase is because crystals growing on the vertical filament
above the $-5$ C region (that is, at higher temperatures) shield the
supersaturation, whereas no crystals are growing above the wire tip. With 2
kV applied, c-axis electric needles grow with tip velocities of 150-200 $\mu 
$m/sec. An analysis of diffusion-limited growth indicates that these
fast-growing electric needles have tip radii of approximately 100 nm during
growth \cite{eneedles}.

\subsection{Diffusion Chamber 2}

DC2 was designed to provide a stable, well-defined ice growth environment
that can also be adjusted over a broad range in temperature and
supersaturation. In particular, the chamber was designed to eliminate
convective air currents as much as possible, so the supersaturation can be
accurately calculated from the known boundary conditions inside the chamber.

To eliminate convection, the isothermal surfaces within the chamber must be
horizontal, which is a nontrivial design feature. Referring to Figure \ref%
{DCschematic1}, DC2 begins with an aluminum clam-shell construction, the top
half being held at a temperature $T_{DC2,top}$ and the bottom half at $%
T_{DC2,bottom}.$ Polycarbonate plates span the gap between the upper and
lower halves, thus sealing the chamber to prevent air currents coming in
from outside. Within the clamshell, 1.6-mm thick stainless steel plates form
a set of inner walls, and their purpose is to provide a uniform linear
temperature gradient along the walls. Heat flow down these plates is
substantially higher than heat flow from the surrounding clamshell (the
latter impeded by the low thermal conductivity of air), thus ensuring the
linear temperature gradient. The inner dimensions of DC2 are
184.1x184.1x101.6 mm.

\begin{figure}[ht] 
  \centering
  \includegraphics[width=4in,keepaspectratio]{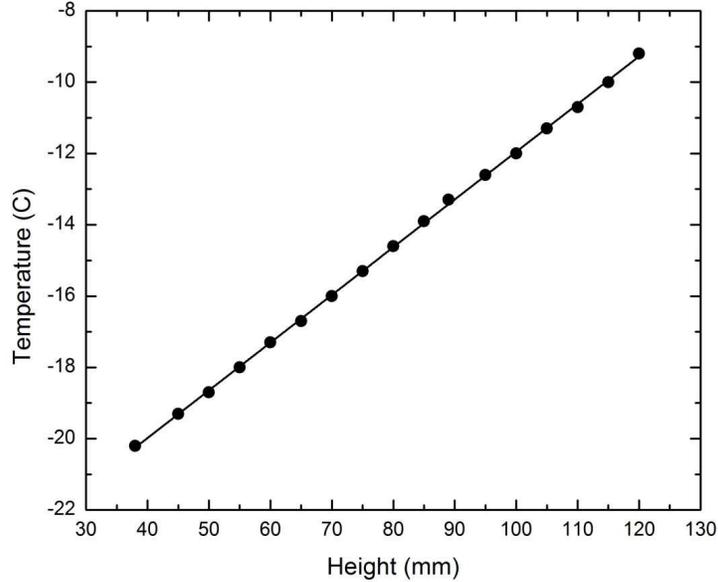}
  \caption{A typical measured vertical
temperature profile along the center axis of the second diffusion chamber
(DC2). The linear profile ensures that the isotermal surfaces within DC2 are
horizonal, suppressing convective currents. In the absence of convection,
the supersaturation can be modeled by solving the diffusion equation inside
the chamber with the appropriate boundary conditions.
A typical measured vertical
temperature profile along the center axis of the second diffusion chamber
(DC2). The linear profile ensures that the isotermal surfaces within DC2 are
horizonal, suppressing convective currents. In the absence of convection,
the supersaturation can be modeled by solving the diffusion equation inside
the chamber with the appropriate boundary conditions.
}
  \label{DC2tempprofile}
\end{figure}

Figure \ref{DC2tempprofile} shows an example of a measured vertical
temperature profile within DC2. For this measurement the top and bottom
plate temperatures were set to $-8$ C and $-22$ C, respectively, giving a
center temperature of $-15$ C. A fit to these data yields a temperature
gradient 0.134 C/mm, which differs from the expected value of 0.138 C/mm
(for a plate separation of 101.6 mm) by three percent. The difference
appears to be caused by slight heat leaks from the side windows in DC2 (each
consisting of three insulated layers between the inner walls and the outside
air), together with the fact that the thermistors used in DC2 are only rated
to an absolute accuracy of 0.1 C.

Solving the heat diffusion equation within DC2 yields the solution 
\begin{eqnarray}
T(z)=T_{DC2,bottom}+(z/\Delta z)(T_{DC2,top}-T_{DC2,bottom})
\end{eqnarray} where this
describes the wall temperature, the top and bottom temperatures, and the air
temperature within the chamber. As long as $T_{DC2,top}>T_{DC2,bottom},$ the
isothermal surfaces will be horizontal and no convection will occur.

The wire assembly passes through a small linear opening at the bottom of
DC2, and this opening is blocked with a slit mylar sheet to reduce air
currents. In addition, the temperatures below the bottom surface of DC2 are
typically lower than $T_{DC2,bottom}$ (see Figure \ref{DCschematic1})
providing a stable temperature gradient that prevents convection through the
opening. The wire also passes through an opening in one wall of the chamber
during its transfer from DC1, and a sliding shutter seals this opening
except for the few seconds during transfer.

With convection essentially eliminated, the particle diffusion equation can
be solved to determine the supersaturation within the chamber. If the walls
were infinitely far away, then the water vapor concentration in the chamber
would be $c(z)=c_{bottom}+(z/\Delta z)(c_{top}-c_{bottom})$ (ignoring small
corrections from, for example, the temperature dependence of the diffusion
constant) and the supersaturation at the center of the chamber would be%
\begin{eqnarray}
\sigma _{center} &\approx &\frac{1}{2}\frac{1}{c_{sat}(T_{center})}\frac{%
d^{2}c_{sat}}{dT^{2}}(T_{center})\left( \Delta T\right) ^{2}  \label{sat} \\
&\approx &0.0032\left( \Delta T\right) ^{2}  \nonumber
\end{eqnarray}%
to lowest order in $\Delta T=(T_{DC2,top}-T_{DC2,bottom})/2,$ where $\Delta T
$ is measured in degrees Celsius in the second line of this equation. Note
that $c_{sat}^{-1}\left( d^{2}c_{sat}/dT^{2}\right) $ is only a weak
function of temperature, so this approximation is adequate for making rough
calculations over the range $0>T>-40$ C.

For better accuracy we numerically solve the diffusion equation within DC2,
including the boundary condition that $c=c_{sat}(T)$ on the ice-covered
walls of the chamber, where $T$ is the surface temperature at each point.
These numerical models indicate that the center supersaturation is
approximately 0.8 times the result given by Equation \ref{sat}. The
ice-covered wire adds an additional correction factor of approximately 0.9,
depending on how much ice is on the wire, the length of the ice needle being
tested, and the locations of any neighboring needles. A careful examination
of these effects gives an absolute fractional uncertainty in the
supersaturation of approximately $\pm 10$ percent under typical operating
conditions. For $T=-15$ C, for example, we find that the center
supersaturation is well described by $\sigma _{center}\approx 0.0023\left(
\Delta T\right) ^{2}$. A more detailed analysis of the supersaturation
within DC2, along with test measurements, will be reported separately.

While DC1 includes an ample water reservoir (100 ml) at the top of the
chamber, DC2 contains four shallow water reservoirs (15 ml each) at the
bottom of the chamber. During the initial cooldown of the system, DC2 is run
in reverse -- with the top cooler than the bottom -- in order to promote
convection that carries water from the bottom reservoirs to the top and
walls of the chamber. These surfaces become coated with water droplets
during this phase (which lasts about three hours), and subsequently cooling
the entire chamber to $T<-20$ C for fifteen minutes freezes the droplets.
This process coats essentially the entire inside surface of DC2 with a thin
layer of ice, while maintaining the simple rectangular geometry of the
chamber. The volume of ice transferred is sufficient that all surfaces will
remain ice covered for the duration of an experimental run (typically less
than ten hours).

\section{Initial Observations}

\begin{figure}[p] 
  \centering
  \includegraphics[width=6.0in,keepaspectratio]{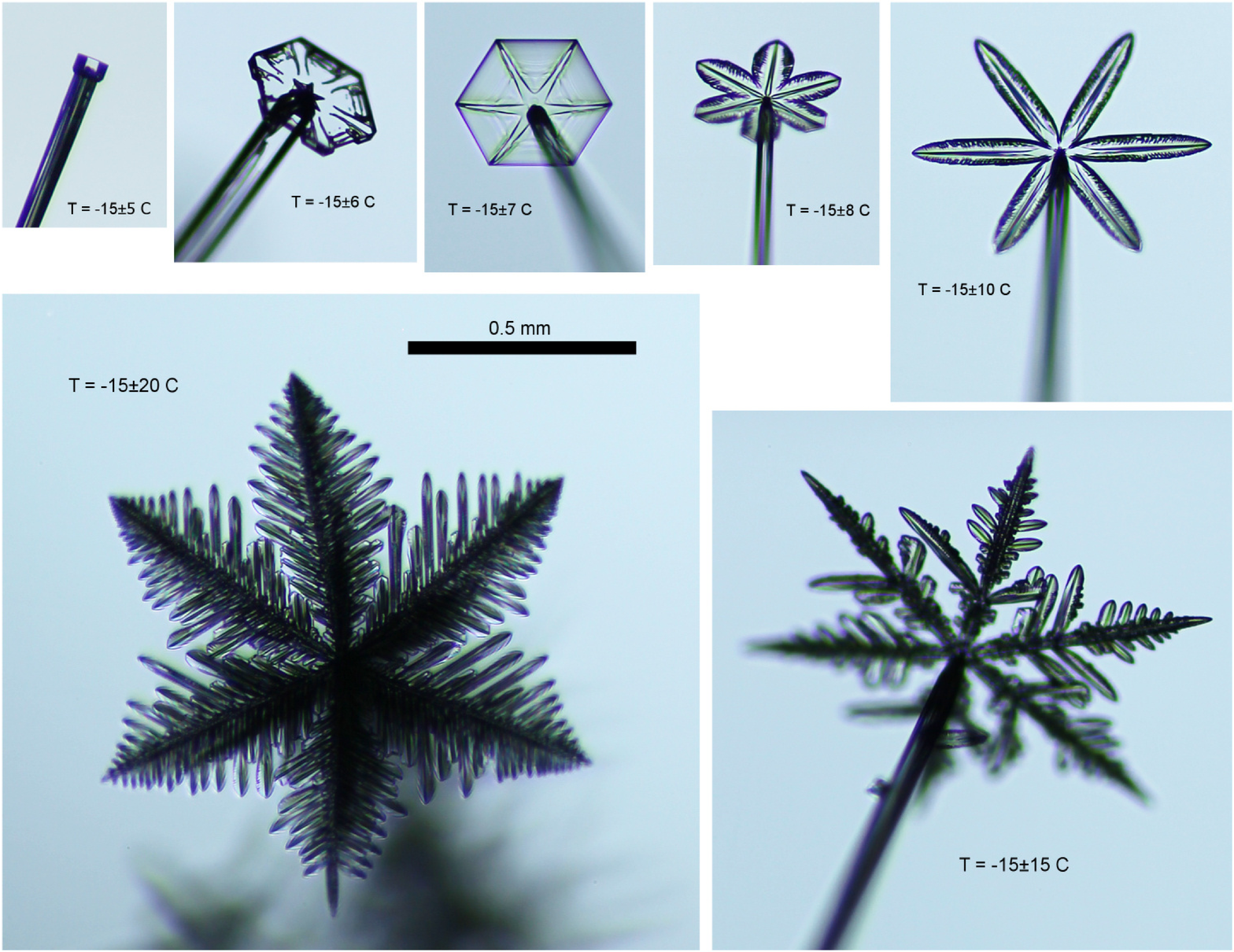}
  \caption{A progression of ice crystals
grown at $-15$ C on the ends of c-axis ice needles, showing the
morphological progression from blocky (upper left) to thick plate, thin
plate, dorite, simple star, stellar dendrite, and fernlike stellar dendrite
as the supersaturation is increased. The images are labeled with $T=T_{0}\pm
\Delta T$ C, where $T_{0}$ is the DC2 center temperature (where the crystals
are growing), $T_{0}+\Delta T$ is the DC2 top surface temperature, and $%
T_{0}-\Delta T$ is the bottom temperature. The images are all shown at the
same scale, so the 0.5 mm scale bar applies to each. The growth times varied
from 13 minutes (blocky) to 1 minute (fernlike stellar dendrite). The
supersaturation seen by the growing crystals is given by $\protect\sigma %
\approx 0.0023(\Delta T)^{2}$, as determined by numerical models of the
water vapor concentration field within DC2.
}
  \label{combo15}
\end{figure}

\begin{figure}[p] 
  \centering
  \includegraphics[width=6.0in, keepaspectratio]{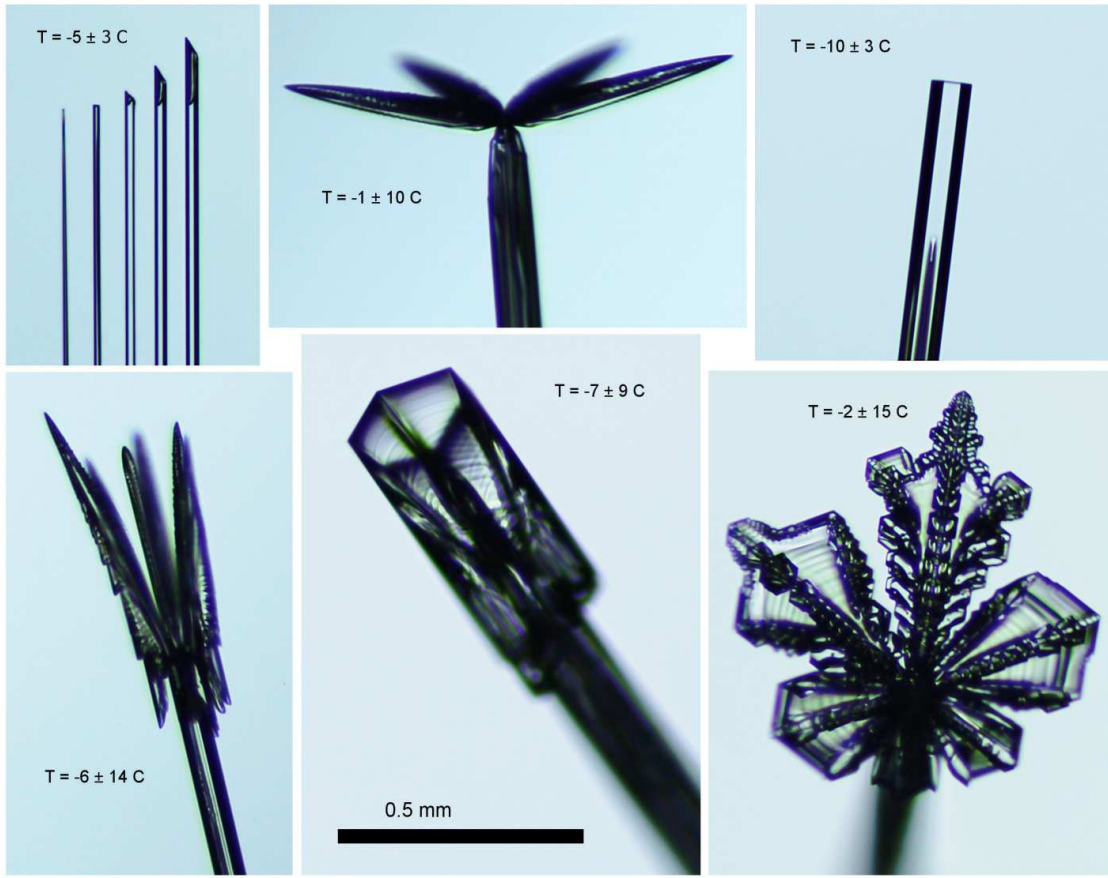}
  \caption{Additional examples of ice
crystals grown in DC2 on the ends of c-axis ice needles. Again the images
are labeled with $T=-T_{0}\pm \Delta T$ C, where $T_{0}$ is the temperature
at the growing crystal, $T_{0}+\Delta T$ is the DC2 top temperature and $%
T_{0}-\Delta T$ is the DC2 bottom temperature. The images are all shown at
the same scale, so the 0.5 mm scale bar applies to each. The composite image
at the upper left combines several images taken at different times, showing
the formation of a thin hollow columnar crystal.
}
  \label{combo2}
\end{figure}

Figure \ref{combo15} shows some example crystals grown on c-axis ice needles
at $-15$ C using this apparatus, showing the progression of morphologies at
this temperature as a function of supersaturation. Figure \ref{combo2} shows
some additional example crystals grown at other temperatures. As described
above, the dual-chamber apparatus is especially well suited to making
quantitative measurements at known temperatures and supersaturations, for
comparison with numerical models of faceted dendritic growth. Figure \ref%
{modeling} shows one of our early attempts at 2D modeling of thin plates
growing on ice needles, using the cellular automata technique described in 
\cite{modeling}. Moving forward to full 3D numerical models remains a
challenge, but recent progress in producing realistic models using accurate
input physics appears quite promising \cite{kelly}.

\begin{figure}[p] 
  \centering
  \includegraphics[width=3.7in,keepaspectratio]{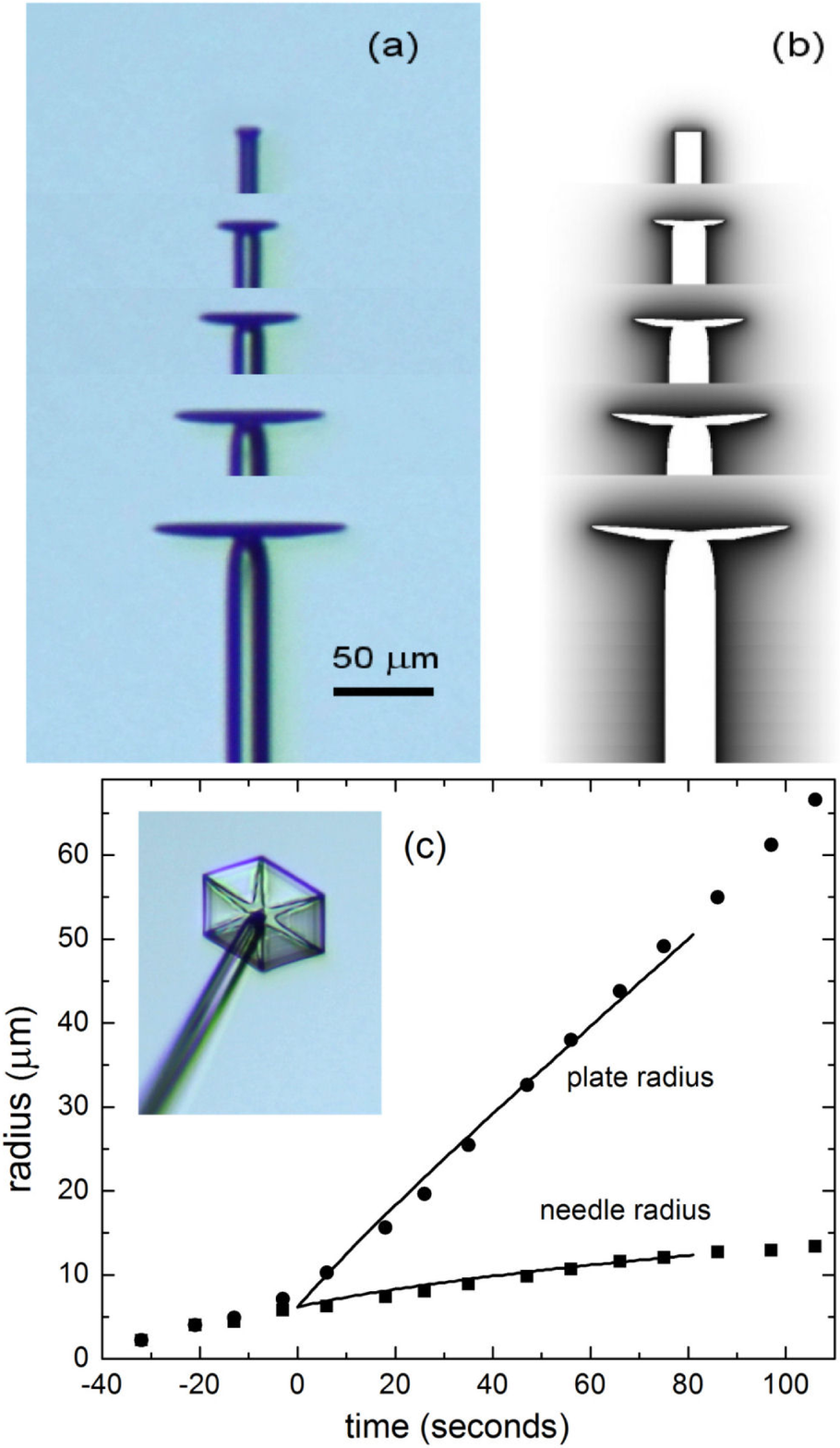}
  \caption{An example showing numerical
modeling of an ice crystal grown at $T=-15$ C and $\protect\sigma =13$
percent, from \protect\cite{modeling}. The composite image in (a) shows five
successive views of a thin plate-like crystal growing on the end of a
slender ice needle, with a 50-$\protect\mu $m scale bar. The crystal is
shown in a side view, with illumination from behind. The composite image in
(b) shows our numerical model of the same crystal. Here the brightness
around the crystal is proportional to supersaturation. The data points in
(c) show measurements of the plate and needle radii (the latter at a
distance of 50 $\protect\mu $m from the base of the plate) as a function of
time. The lines in the graph are from the growth of the model crystal. The
inset image in (c) shows a more frontal view of a similar plate-on-needle
crystal with a plate radius of 80 $\protect\mu $m, showing the thin
sectored-plate morphology. Note that at earlier times $(t<0$ in the graph)
the crystal was growing from a tapered needle to a nearly uniform column,
from which the plate emerged. The time axes were shifted so the plate began
growing at $t\approx 0$ for both the real and model crystals.
}
  \label{modeling}
\end{figure}


\begin{thebibliography}{99}
\bibitem{griffeathgravner} J. Gravner and D. Griffeath, \textquotedblleft
Modeling snow-crystal growth: A three-dimensional mesoscopic
approach,\textquotedblright\ Phys. Rev. E 79, 011601 (2009).

\bibitem{barrettgarcke} J. W. Barrett, H. Garcke, and R. Nurnberg,
\textquotedblleft Numerical computations of faceted pattern formation in
snow crystal growth,\textquotedblright\ Phys. Rev. E 86, 011604 (2012).

\bibitem{libbrechtreview} K. G. Libbrecht, \textquotedblleft The physics of
snow crystals,\textquotedblright\ Rep. Prog. Phys., 68, 855-895 (2005).

\bibitem{comprehensive} K. G. Libbrecht, \textquotedblleft Toward a
comprehensive model of snow crystal growth dynamics: 1. Overarching features
and physical origins,\textquotedblright\ arXiv:1211.5555 (2012).

\bibitem{critical} K. G. Libbrecht, \textquotedblleft A critical look at ice
crystal growth data,\textquotedblright\ arXiv:cond-mat/0411662 (2004).

\bibitem{growthmeas} K. G. Libbrecht and M. E. Rickerby, \textquotedblleft
Measurements of surface attachment kinetics for faceted ice crystal
growth,\textquotedblright\ J. Cryst. Growth 377, 1-8 (2013).

\bibitem{systematicerrors} K. G. Libbrecht, \textquotedblleft Managing
systematic errors in ice crystal growth experiments,\textquotedblright\
arXiv:1208.5064 (2012).

\bibitem{nakaya} Nakaya, U., \textquotedblleft Snow Crystals: Natural and
Artificial,\textquotedblright\ (Harvard University Press) (1954).

\bibitem{knight5} C. A. Knight, \textquotedblleft Ice growth from the vapor
at -5 degrees C,\textquotedblright\ J. Atmo. Sci. 69, 2031-40 (2012).

\bibitem{knightnelson} J. Nelson and C. Knight, \textquotedblleft Snow
crystal habit changes explained by layer nucleation,\textquotedblright\ J.
Atmo. Sci. 55, 1452-65 (1998).

\bibitem{gonda} T. Gonda and M. Komabayasi, \textquotedblleft Skeletal and
dendritic structures of ice crystal as a function of thermal conductivity
and vapor diffusivity,\textquotedblright\ J. Meteor. Soc. Japan 49, 32--41
(1971).

\bibitem{gonda2} T. Gonda, \textquotedblleft The growth of small ice
crystals in gases of high and low pressures,\textquotedblright\ J. Meteor.
Soc. Japan 54, 233--40 (1976).

\bibitem{triangular} K. G. Libbrecht and H. M. Arnold, \textquotedblleft
Aerodynamic stability and the growth of triangular snow
crystals,\textquotedblright\ arXiv:0911.4267 (2009).

\bibitem{levitated} B. D. Swanson, M. J. Bacon, E. J. Davis, and M. B.
Baker, \textquotedblleft Electrodynamic trapping and manipulation of ice
crystals,\textquotedblright\ Quart. J. Royal Meteor. Soc. 125, 1039-58
(1999).

\bibitem{fukuta} T. Takahasi, T. Endoh, G. Wakahama, and N. Fukuta,
\textquotedblleft Vapor diffusional growth of free-falling snow
crystals,\textquotedblright\ J. Meteor. Soc. Japan 69, 15--30 (1991).

\bibitem{levitatetherm} T. Kelling, G. Wurm, and C. Durmann,
\textquotedblleft Ice particles trapped by temperature gradients at mbar
pressures,\textquotedblright\ Rev. Sci. Instr. 82, 115105 (2011).

\bibitem{acoustic} Y. J. Lu, W. J. Xie, and B. Wei, \textquotedblleft
Observation of ice nucleation in acoustically levitated water
drops,\textquotedblright\ Appl. Phys. Lett. 87, 184107 (2005).

\bibitem{eneedles} K. G. Libbrecht, T. Crosby, and M. Swanson,
\textquotedblleft Electrically enhanced free dendrite growth in polar and
non-polar systems,\textquotedblright\ J. Cryst. Growth 240, 241-9 (2002).

\bibitem{fishbones} K. G. Libbrecht, \textquotedblleft Identification of a
novel "fishbone" structure in the dendritic growth of columnar ice
crystals,\textquotedblright\ arXiv:0912.2522 (2009).

\bibitem{modeling} K. G. Libbrecht, \textquotedblleft Quantitative modeling
of faceted ice crystal growth from water vapor using cellular
automata,\textquotedblright\ J. Comp. Meth. Phys. 174806 (2013).

\bibitem{kelly} J. G. Kelly and E. C. Boyer, \textquotedblleft Physical
improvements to a mesoscopic cellular automaton model for three-dimensional
snow crystal growth, arXiv:1308.4910 (2013).
\end{thebibliography}
\end{document}